\documentclass[12pt]{article}

\def \ov {\over}

\newcounter{subequation}[equation]

\newcommand{\be}{\begin{equation}}
\newcommand{\ee}{\end{equation}}
\newcommand{\eel}[1]{\label{#1}\end{equation}}
\newcommand{\bea}{\begin{eqnarray}}
\newcommand{\eea}{\end{eqnarray}} 
\newcommand{\eeal}[1]{\label{#1}\end{eqnarray}}

\makeatletter

\def\thesubequation{\theequation\@alph\c@subequation}
\def\@subeqnnum{{\rm (\thesubequation)}}
\def\slabel#1{\@bsphack\if@filesw {\let\thepage\relax
   \xdef\@gtempa{\write\@auxout{\string
      \newlabel{#1}{{\thesubequation}{\thepage}}}}}\@gtempa
   \if@nobreak \ifvmode\nobreak\fi\fi\fi\@esphack}
\def\subeqnarray{\stepcounter{equation}
\let\@currentlabel=\theequation\global\c@subequation\@ne
\global\@eqnswtrue \global\@eqcnt\z@\tabskip\@centering\let\\=\@subeqncr

$$\halign to \displaywidth\bgroup\@eqnsel\hskip\@centering
  $\displaystyle\tabskip\z@{##}$&\global\@eqcnt\@ne
  \hskip 2\arraycolsep \hfil${##}$\hfil
  &\global\@eqcnt\tw@ \hskip 2\arraycolsep
  $\displaystyle\tabskip\z@{##}$\hfil
   \tabskip\@centering&\llap{##}\tabskip\z@\cr}
\def\endsubeqnarray{\@@subeqncr\egroup
                     $$\global\@ignoretrue}
\def\@subeqncr{{\ifnum0=`}\fi\@ifstar{\global\@eqpen\@M
    \@ysubeqncr}{\global\@eqpen\interdisplaylinepenalty \@ysubeqncr}}
\def\@ysubeqncr{\@ifnextchar [{\@xsubeqncr}{\@xsubeqncr[\z@]}}
\def\@xsubeqncr[#1]{\ifnum0=`{\fi}\@@subeqncr
   \noalign{\penalty\@eqpen\vskip\jot\vskip #1\relax}}
\def\@@subeqncr{\let\@tempa\relax
    \ifcase\@eqcnt \def\@tempa{& & &}\or \def\@tempa{& &}
      \else \def\@tempa{&}\fi
     \@tempa \if@eqnsw\@subeqnnum\refstepcounter{subequation}\fi
     \global\@eqnswtrue\global\@eqcnt\z@\cr}
\let\@ssubeqncr=\@subeqncr
\@namedef{subeqnarray*}{\def\@subeqncr{\nonumber\@ssubeqncr}\subeqnarray}

\@namedef{endsubeqnarray*}{\global\advance\c@equation\m@ne%
                           \nonumber\endsubeqnarray}

\makeatletter \@addtoreset{equation}{section} \makeatother
\renewcommand{\theequation}{\thesection.\arabic{equation}}

\catcode`\@=11

\newcount\hour
\newcount\minute
\newtoks\amorpm \hour=\time\divide\hour by 60\minute
=\time{\multiply\hour by 60 \global\advance\minute by-\hour}
\edef\standardtime{{\ifnum\hour<12 \global\amorpm={am}%
        \else\global\amorpm={pm}\advance\hour by-12 \fi
        \ifnum\hour=0 \hour=12 \fi
        \number\hour:\ifnum\minute<10
        0\fi\number\minute\the\amorpm}}
\edef\militarytime{\number\hour:\ifnum\minute<10 0\fi\number\minute}

\def\draftlabel#1{{\@bsphack\if@filesw {\let\thepage\relax
   \xdef\@gtempa{\write\@auxout{\string
      \newlabel{#1}{{\@currentlabel}{\thepage}}}}}\@gtempa
   \if@nobreak \ifvmode\nobreak\fi\fi\fi\@esphack}
        \gdef\@eqnlabel{#1}}
\def\@eqnlabel{}
\def\@vacuum{}
\def\marginnote#1{}
\def\draftmarginnote#1{\marginpar{\raggedright\scriptsize\tt#1}}
\def\draft{
        \pagestyle{plain}
        \overfullrule=2pt
        \oddsidemargin -.5truein
        \def\@oddhead{\sl \phantom{\today\quad\militarytime} \hfil
        \smash{\Large\sl DRAFT} \hfil \today\quad\militarytime}
        \let\@evenhead\@oddhead
        \let\label=\draftlabel
        \let\marginnote=\draftmarginnote
        \def\ps@empty{\let\@mkboth\@gobbletwo
        \def\@oddfoot{\hfil \smash{\Large\sl DRAFT} \hfil}
        \let\@evenfoot\@oddhead}

\def\@eqnnum{(\theequation)\rlap{\kern\marginparsep\tt\@eqnlabel}%
        \global\let\@eqnlabel\@vacuum}  }

\renewcommand{\theequation}{\thesection.\arabic{equation}}
\renewcommand{\thefootnote}{\fnsymbol{footnote}}

\def\appendix#1{
  \addtocounter{section}{-3}
  \setcounter{equation}{0}
  \renewcommand{\thesection}{\Alph{section}}
  \section*{Appendix \thesection\protect\indent \parbox[t]{11.15cm}
  {#1} }
  \addcontentsline{toc}{section}{Appendix \thesection\ \ \ #1}
  }
\def \ov {\over}

\def\be{\begin{equation}}
\def\ee{\end{equation}}

\date{}
\begin{document}

\begin{titlepage}



\begin{center}

{\Large \bf  One Loop Partition Function in Plane Waves R-R Background}

\vskip .7 cm

\vskip 1 cm

{\large   Amine B. Hammou}\\

\end{center}

\vskip 0.4cm \centerline{\it Department of Physics and Institute of
Plasma Physics,} 
\vskip 0.2 cm 
\centerline{\it University of Crete, and FO.R.T.H. 71003 Heraklion GREECE}

\vskip 3 cm

\begin{abstract}
We compute the one loop partition function of type IIB string in plane wave
R-R 5-form background $F^5$ using both path integral and operator formalisms 
and show that the two results agree perfectly. The result turns out to be 
equal to the partition function in the flat background. We also study the 
Tadpole cancellation for the 
unoriented closed and open string model in plane wave R-R 5-form 
background studied in hep-th/0203249 and find that the cancellation of the 
Tadpole requires the gauge group to be SO(8).  
\end{abstract}


\vskip 5cm  
\mbox{~~~~~~~email address: amine@physics.uoc.gr}

\end{titlepage}
\setcounter{page}{1} \renewcommand{\thefootnote}{\arabic{footnote}}
\setcounter{footnote}{0}

\def \N{{\cal N}} \def \ov {\over}

\setcounter{page}{1} \renewcommand{\thefootnote}{\arabic{footnote}}
\setcounter{footnote}{0}

\def \N{{\cal N}} \def \ov {\over}

\section{Introduction}
 
Recently a new maximally supersymmetric solution of type IIB supergravity
was found \cite{blau1} and it was related by a special limit 
(Penrose limit) to the famous $AdS_5\times S^5$ \cite{blau2}. It turns out 
to be a ten dimensional plane wave space with metric and R-R 5-form flux

\bea
ds^2&=&-2dx^+ dx^- -\mu^2 x_i x^i(dx^+)^2+dx_i dx^i,\nonumber\\
F_{+1234}&=&F_{+5678}=2\mu.
\label{ppwave}
\eea

Remarkably, the Green-Schwarz formulation of the theory shows up to be 
exactly solvable in the light-cone gauge \cite{m,rt}. Moreover, the exact
string spectrum in plane wave has been successfully compared to the dual
${\cal N}=4$ gauge theory operators \cite{bmn}. A similar comparison 
has been  carried out in \cite{gnbmn} for a plane wave model dual
to an ${\cal N}=2$ $Sp(N)$ gauge theory with a hypermultiplet in the 
antisymmetric representation and four fundamental hypermultiplets.
This theory is dual to string theory on $AdS_5\times S^5/Z_2$,
where the $Z_2$ is an orientifold action \cite{FS}. In this paper we
will consider this model and study its anomaly cancellation -Tadpoles-, 
this is important for the consistency of the theory, and to fix the gauge 
group which was crucial in the comparison with the 
dual ${\cal N}=2$ gauge theory \cite{gnbmn}.    
The paper is organized as follows. 
Section 2 we briefly review the light cone gauge world sheet theory
of type IIB strings in plane wave R-R background. In section 3 we compute
the partition function using both path integral and operator formalisms and
show that the two results agree and are equal to the partition 
function in flat space background. Finally in section 4 we will study the 
anomaly cancellation -Tadpoles- for a model of unoriented closed and open 
strings in plane wave R-R background considered in \cite{gnbmn}.   

\section{Review of type IIB in plane wave R-R background}

The $\kappa$-symmetry gauge fixed type IIB GS Lagrangian in the plane wave 
R-R background i.e. 
$\bar{\gamma}^{+}\theta=\bar{\gamma}^{+}\bar{\theta}=0$ is given by \cite{m}:

\begin{eqnarray}\label{boson}
{\cal L}_{B} & = &-\frac{1}{2}\sqrt{g}g^{ab}\Bigl(2\partial_a
x^+
\partial_b x^- -x_i^2 \partial_a x^+\partial_b x^+ + \partial_a
x^i \partial_b x^i\Bigr)
\end{eqnarray}

\begin{eqnarray}
\label{fermion} 
{\cal L}_F &=& -  {\rm i}\sqrt{g}g^{ab}\partial_b x^+\Bigl(
\bar{\theta}\bar{\gamma}^-\partial_a\theta + \theta\bar{\gamma}^-
\partial_a\bar{\theta} + 2{\rm i}\partial_a x^+
\bar{\theta}\bar{\gamma}^- \Pi \theta\Bigr) \nonumber\\
&+& \epsilon^{ab}\partial_a x^+ (\theta
\bar{\gamma}^-\partial_b \theta +\bar{\theta}
\bar{\gamma}^-\partial_b \bar{\theta})\,.\end{eqnarray}

where $i=1,...,8$ and 
$\Pi =\gamma^{1}\bar{\gamma}^{2}\gamma^{3}\bar{\gamma}^4$. As in flat 
space case we can 
eliminate the $\partial x^{+}$-factors from the kinetic 
terms of the fermionic fields. This is possible by choosing the bosonic 
light-cone 
gauge which fixes the residual conformal symmetry

\be
\sqrt{g} g^{ab} = \eta^{ab}~,~~~~~~~~~~~~~~~~~~x^{+}(\tau,\sigma)=\alpha' 
p^{+} \tau.
\ee

The resulting Lagrangian will be that of $8$ free massive two dimensional 
bosons and $8$ free massive 
Majorana two dimensional fermionic fields propagating in a flat two 
dimensional space with mass 
$m= \alpha' \mu p^+$. Note that we can recover the flat case in the limit 
$\mu\to 0$, while the limit 
$\alpha' p^+ \to 0$ corresponds to supergravity in the plane wave 
background \cite{m}.

The Light-cone superstring Hamiltonian can be expressed as\footnote{We 
will put $\alpha'=1$ in what follows, then all one 
needs to do to restore the dependence on the string tension is to rescale
$p^{+}\to \alpha' p^+$ \cite{m}.}:

\bea
H &=& \mu(a_0^i {}^\dagger a_0^i + S_0^a {}^\dagger S_0^a)\nonumber\\
&&+\frac{1}{p^+} \sum_{n=1}^\infty \omega_n \left(
a_n^i {}^\dagger a_n^i + \tilde{a}_n^i {}^\dagger \tilde{a}_n^i
+S_n^a {}^\dagger S_n^a+\tilde{S}_n^a {}^\dagger \tilde{S}_n^a\right),
\label{hcl}
\eea
   
where 

\be
\omega_n=\sqrt{m^2+n^2},
\nonumber
\ee

and we are summing over the repeated indices $i$ and $a$ going from 
$1,...,8$.  
The bosonic zero modes are defined through  

\be 
\label{bzm} 
a_0^{i} = \frac{1}{\sqrt{2 m}}(p_0^i + i m x_0^i),\,\,\,\, 
a_0^i{}^+ =\frac{1}{\sqrt{2 m}}( p_0^i - i m x_0^i)\,, 
\ee

$p_0^i$ and $x_0^i$ are the zero modes of the bosonic fields, which in 
the flat limit
will correspond to the momentum and the center of mass positions. The fermionic
fields $S$ and $\tilde{S}$ are related to the fields $\theta$ and 
$\bar{\theta}$ in the same way
as in the flat space. The vacuum state
is defined as the direct product of the zero mode vacuum and the vacuum of 
the string oscillation modes annihilated by all the annihilation operators 
appearing in the Hamiltonian
(\ref{hcl}). Generic Fock space states are then obtained by acting on the 
vacuum with the creation
operators and the subspace of physical states is obtained by imposing the 
level matching condition
i.e. $N = \tilde{N}$ with

\be
N =\sum_{n=1}^\infty n (a_n^i {}^\dagger a_n^i
+S_n^a {}^\dagger S_n^a),
\label{number}
\ee

and the same expression for $\tilde{N}$ with tilde on the oscillators.

\section{Closed String Partition function}

In this section we will compute the partition function of type IIB 
superstring in plane wave 
R-R background described above, using the path integral formalism 
\footnote{I would like to thank 
very much A.A. Tseytlin for correspondence on this point \cite{tnw}.} and 
show that the result is zero as it 
should be for any globally supersymmetric background. Surprisingly, it turns 
out that the 
partition function of type IIB in plane wave R-R background is equal to 
the partition function of 
type IIB in flat background. This result seems to contradict the result 
obtained in \cite{takayanagi} using the 
operator formalism. We will show that in the operator formalism the partition 
function computed by 
Takayanagi \cite{takayanagi} is also zero after a careful integration over 
$p^+$ and $p^-$. This is to be expected 
since the two results can be compared only after performing the integral over 
the momentum.

\subsection{\it Path Integral}

Following Kallosh and Morozov \cite{KM} \footnote{I am very grateful to 
A.A. Tseytlin for drawing my attention to this paper.}
the partition function in the 
path integral formalism can be computed by fixing
the semi-light cone gauge i.e. 
$\bar{\gamma}^+ \theta =\bar{\gamma}^+ \bar{\theta}= 0$ and 
$g_{ab} = \rho g_{ab}^{(m)}$, where
$g_{ab}^{(m)}$ is some background metric.  In 
conformal coordinates $z$, 
$\bar{z}$ with $g_{zz}^{(m)}=g_{\bar{z}\bar{z}}^{(m)}=0$, 
$g_{z\bar{z}}^{(m)}=\frac{1}{2}$ omitting 
the integration over the background moduli, the gauge fixed path integral 
will take the form 

\be
\int Dx^\mu D\bar{\theta} D\theta Db Dc \,\,\,(Det u_{\bar{z}})^{-8} 
e^{-S-S_{gh}},
\label{path}
\ee

where $u_{\bar{z}}=\partial_{\bar{z}} x^+$ and the action $S$ 
stand for the covariant action associated to the bosonic and fermionic 
Lagrangian (\ref{boson})-(\ref{fermion}) expressed 
in conformal coordinates $z$, $\bar{z}$    
    
\begin{eqnarray}\label{action}
S &=& \int d^2 z \Bigl[2\partial_z
x^{\mu} \partial_{\bar{z}} x_{\mu}-x_i^2 \partial_z x^+
\partial_{\bar{z}} x^+ \nonumber\\
&+& {\rm i}  \partial_{\bar{z}} x^+ 
(\theta^1 \bar{\gamma}^-\partial_z \theta^1 + \theta^2 \bar{\gamma}^-
\partial_z \theta^2 + 4 {\rm i}\partial_z x^+
\theta^1 \bar{\gamma}^- \Pi \theta^2 )\Bigr],
\end{eqnarray}
 
where $\mu=+,-,i$ and $\theta^1=\frac{1}{\sqrt{2}}(\theta+\bar{\theta})$,
$\theta^2=-\frac{i}{\sqrt{2}}(\theta-\bar{\theta})$ are two real 
Majorana-Weyl spinors which would correspond to the left and right moving 
fermions in the flat background limit. The ghosts action is as usual given by

\be
S_{gh}=\int d^2 z (b \partial_{\bar{z}} c +\bar{b} \partial_z \bar{c}).
\ee

The $(Det u_{\bar{z}})^8$ is the local measure of the integration related to 
the 
Faddeev-Popov fermionic gauge symmetry nonpropagating ghosts. It provides 
the independence
of the theory of the way in which local fermionic gauge symmetry, 
$\kappa$-symmetry, was fixed. Using the
fact that $SO(8)$ spinors ($\bar{\gamma}^+ \theta^{\cal I}=0$) may be 
presented as two $SU(4)$ spinors 
($\eta_{\cal I}^k$, $\theta^{\cal I}_k$), $k=1,...,4$ and ${\cal I}=1,2$. 
Then 
$\theta^{\cal I}\gamma^{-} u_{\bar{z}} \partial_{z} \theta^{\cal I}\to 
\eta_{\cal I}^k 
u_{\bar{z}}
\partial_{z} \theta^{\cal I}_k +\frac{1}{2}\eta_{\cal I}^k \theta^{\cal I}_k
\partial_z u_{\bar{z}}$. The last term $\eta_{\cal I}^k\theta^{\cal I}_k
\partial_z \partial_{\bar{z}} x^+$ can be absorbed into a redefinition of 
$x^-$. We can absorb the factor $u_{\bar{z}}$ in a redefinition of the spinors 
$\eta_{\bar{z},{\cal I}}^k=\eta_{\cal I}^k \partial_{\bar{z}} x^+$, 
then the Jacobian of this transformation will cancel the factor 
$(Det u_{\bar{z}})^8$ appearing in the measure (\ref{path}). We can now 
perform the integration over $x^-$ which will give a delta 
function\footnote{We are considering a Euclidean space time by means of a wick 
rotation, hence we should integrate our final result over a Euclidean torus
\cite{CKR}.}     
$\delta(\partial_z\partial_{\bar{z}} x^+)$, and over $x^+$ which will first 
constrain 
$x^+$ to be constant on the Torus i.e. $\partial_z\partial_{\bar{z}} x^{+}=0$
and second cancel the contribution of the ghosts 
to the path integral\footnote{as usual the contribution of the ghost to the
path integral will cancel the same contribution coming from the bosonic 
fields in the light-cone directions.}. Therefore, all the terms containing
$\partial_z x^+$ in the action will cancel. These are 
the terms which was 
the source for the mass in the bosonic light-cone gauge theory discussed in
the previous section. The path integral will become

\be
\int Dx^{i} D\eta_{\bar{z}, {\cal I}}^k D\theta^{\cal I}_k \,\,\,
exp \left\{-\int d^2 z\,\,\,(\partial_{z} x^i \partial_{\bar{z}} x^i
+\eta_{\bar{z},{\cal I}}^k\partial_z \theta^{\cal I}_k)\right\},
\nonumber\\
\ee

which is nothing but the path integral of the bosonic and fermionic fields
in the transverse directions. Therefore, the partition function is zero (as it
should be due to supersymmetry) due to
fermionic zero modes \cite{tnw} and it is equal to the partition function of
type IIB string theory in flat background.  
     
\subsection{\it Operator Formalism}

Now we would like to do the same computation using the Operator formalism, 
and show that it leads
to the same result. Following Takayanagi \cite{takayanagi} we define the 
partition function as  

\be
Z = \int_{\cal F} \frac{d\tau_1 d\tau_2}{\tau_2}\int dp^+ dp^- 
\,\, \mbox{Tr} \,\,(-1)^F
e^{-2\pi \tau_2 p^+ (p^- +H)+2\pi i\tau_1 (N-\tilde{N})},
\label{torus}
\ee

where $H$ is the light-cone Hamiltonian (\ref{hcl}), $N$ and $\tilde{N}$ 
are the number 
operators (\ref{number}). $\mbox{Tr}$ denotes the trace in the 
Hilbert space of the 
light-cone string theory and $F$ is the fermion number operator. The modular 
invariance of the 
partition function enables us to restrict the integration of the 
torus moduli $\tau,\bar{\tau}$ to the fundamental region denoted ${\cal F}$.

Performing the trace over the Hilbert space we find

\be
Z_{\mbox{IIB}}=\int_{\cal F}\frac{d\tau_1 d\tau_2}{\tau_2}\int dp^+dp^-
e^{-2\pi\tau_2 p^+ p^-}\left(
\frac{Z_{f}(\tau_1,\tau_2|m)}
{Z_{B}(\tau_1,\tau_2|m)}\right)^4, \label{flatZ}
\ee

where $Z_f$ and $Z_B$ are the partition functions of a massive fermion and 
boson respectively with mass $m=\mu p^+\neq 0$ given by

\bea
Z_{f,B}(\tau,\bar{\tau}|m)&=&e^{4\pi\tau_2\Delta_{(m)}}\,\,\,
(1-e^{-2\pi \tau_2 m})^2
\prod_{n=1}^{\infty}(1-e^{-2\pi\tau_2\sqrt{m^2+n^2}
+2\pi i\tau_1 n})^2\nonumber\\
&&\, \, \, \, \, \, \, \times (1-e^{-2\pi\tau_2\sqrt{m^2+n^2}
-2\pi i\tau_1 n})^2.
\label{ZI}
\eea

The factor $\Delta_{(m)}$ corresponds to the zero point energy 
defined by

\be
\Delta_{(m)}=-\frac{1}{2\pi^2}\sum_{p=1}^{\infty}\int^{\infty}_{0} ds 
\, \, e^{-p^2s-\frac{\pi^2m^2}{s}},
\label{zeropoint}
\ee
and in the massless limit it reproduces the known value $-1/12$.
It is easy to show that (\ref{flatZ}) is modular invariant provided
we also change the momentum $p^+$ and $p^-$ to $|\tau|p^+$ and $|\tau|p^-$ 
respectively \cite{takayanagi}. 
Finally to compare the result of the partition function in the two formalisms 
we should perform 
the integration over $p^{+}$ and $p^{-}$ and since we are doing the 
computation in the 
Euclidean space they are complex variables. Since the integrand is a 
function of only $p^+$ then 
the integral will be the integrand valued at $p^{+}=0$. 
The functions $Z_f$ and $Z_B$ (\ref{ZI}) are equal if $p^{+}\neq 0$ but in the 
$p^{+}\to 0$ limit
they are different, and the difference comes from the zero mode part.
In fact, the limit $p^{+}\to 0$ of $p^{+}H$ will reproduce exactly the 
partition 
function of type IIB in the flat background. The factor 
$(1-e^{-2\pi \tau_2 m})^8$ in $Z_f$ will goes
to zero reproducing the $(8-8)^2$ of the flat background limit. 
Whereas, in $Z_B$ using (\ref{bzm}) it is easy to see that the trace over the 
bosonic zero mode part in the limit $p^{+}\to 0$ 
will just become an integration over non-compact momentum $p_0^i$ which 
now become continue, reproducing the familiar 
factor of $1/\tau_2^4$. Therefore, we will find

\be
Z_{\mbox{IIB}}=\int_{\cal F}\frac{d\tau_1 d\tau_2}{\tau_2^6}\,\,\,
\frac{\theta_{1}^4(\tau)}
{\eta^{12}(\tau)}\,\,\,\frac{\bar{\theta}_{1}^4(\bar{\tau})}
{\bar{\eta}^{12}(\bar{\tau})}. \label{fZ}
\ee
It is worth noting that the limit $p^{+}\to 0$ of $H$ is not without 
ambiguities but what we have really done is the limit $p^{+}\to 0$ of 
$p^{+}H$ which is well defined and it corresponds to the flat 
background limit in the G-S formulation. Similar observations was 
made in \cite{kiritsis} where the torus partition function of the WZW model
for a non semi-simple group was shown to be equal to the flat space value. 
Following the same arguments as for the R-R 5-form background we can show that 
the partition function in the R-R 3-form backgrounds of \cite{rt} is also 
trivial\footnote{For the case of NS-NS 3-form background it was argued
in \cite{rt} that the partition function vanishes by virtue of supersymmetry
of the plane wave background and is the same as the flat space partition
function.}.

\section{Open string in Plane wave R-R background}

In this section we will start by giving a brief review of the model 
of unoriented closed and open strings in plane wave R-R background considered 
in \cite{gnbmn}
and study its Tadpole cancellation. We will show that the model is 
free of anomalies if the gauge group is $SO(8)$ \footnote{When this work 
was nearing completion there appeared a paper \cite{AN} where the authors 
analysed the Tadpole condition in an orientifold of type IIB string theory 
in the plane wave background supported by null R-R 3-form flux $F^3$, so we 
will some times refer to it.}.

\subsection{Review of open string model}

The model studied in \cite{gnbmn} consist of the $O(7)$ orientifold
projection of the plane wave solution of 10d type IIB supergravity.
Since the $O(7)$ plane carries -4 units of D7-brane charge, we should add 
4 D7 branes to cancel the charge locally. This will produce an 
$SO(8)$ gauge group on the world volume of the D7 branes. 
An other way to obtain this system is to start with 
type IIB theory compactified on $S^1 \times S^1$
with $x^{7,8}$-coordinates, then 
project by the world-sheet parity $\Omega$ which gives type I theory. 
Do two T-dualities on the compact directions $x^{7,8}$, this will change 
D9 branes to D7 branes transverse to the compact directions. 
The resulting theory is type IIB projected out
by $\Omega (-1)^{F_L} R$, where $F_L$ is the left fermion number and $R$ is 
a $Z_2$ projection $x^{7,8}\to - x^{7,8}$. There are four fixed points where
the orientifold planes are siting, so we should add the D7 branes
to cancel locally and globally the total charge. Unless type I theory, 
this model have D3 branes in its BPS spectrum.  
As in type IIB theory we can study the Penrose limit of
the near horizon geometry of a D3 brane in this model (O7-D7-D3)\cite{gnbmn}. 
Since, the orientifold is parallel to the light cone directions, the theory
in this background result in an exactly solvable string theory.   

\vskip 0.5 cm
{\large \it Closed Strings}\\

The closed sector of this model is the same as type IIB theory in 
plane wave R-R background. The light cone Hamiltonian is 
given by (\ref{hcl}). The $Z_2$ projection $\Omega (-1)^{F_L} R$ acts on the 
bosonic and fermionic oscillator modes as 
\bea
&&a_n^{i}\to \tilde{a}_n^{i} \,\,\, i=1,...,6\,\,\,\,\,; a_n^{i}\to 
-\tilde{a}_n^{i}\,\,\, i=7,8 \nonumber\\
&&S_n\to i \gamma^{56}\tilde{S}_n
\label{z2cl}
\eea
where the action of $\gamma^{56}$ will be to  give a minus sign to half of the 
fermionic oscillator modes.

\vskip 0.5 cm
{\large \it Open Strings}\\ 

There are also open strings that are stretched between the $D7$ branes. In 
the Penrose limit under 
consideration these branes are located at the origin of the 78-plane meaning 
that $x^7$ and $x^8$
should satisfy Dirichlet boundary conditions at $\sigma=0,\pi$ 
\bea
&&\partial_{\sigma} x^i=0 \,\,\,\,\,\, i=1,...,6\,\,;\,\,\,\,\,\,\,\,\,\,\,
\,\, x^{i}=0\,\,\,\,\,\, i=7,8
\nonumber\\ 
&&S=\gamma^{78}\tilde{S}
\nonumber
\eea

The light cone open string Hamiltonian is given by

\bea
H =\frac{1}{2}\left(\mu (a_0^I {}^\dagger a_0^I + S_0^A {}^\dagger S_0^A+2)
+\frac{1}{p^+} \sum_{n=1}^\infty \omega_n \left(
a_n^i {}^\dagger a_n^i + S_n^a {}^\dagger S_n^a\right)\right),
\label{hop}
\eea
where as before $i,a=1,...,8$, while $I=1,...,6$ and $A=1,...,4$. Note that 
the difference with respect to the 
closed string case is that here we have 6 Neumann directions $x^I$ which have 
zero modes, and the two Dirichlet directions instead will not have 
zero modes. For the 
fermions there are in the other hand 8 zero modes of which 4 of them are 
creation and 4 annihilation operators, which is reflecting the fact that 
now we have half of the original type IIB supersymmetry. Finally the factor 
of 2 in the zero mode part of the Hamiltonian (\ref{hop}) is the vacuum energy 
due to the mismatch between the bosonic and fermionic zero modes. Finally,
the orientifold $Z_2$ projection on the different bosonic and fermionic
oscillator modes is
 
\begin{equation}
a^{i}_n \rightarrow (-1)^n a^{i}_n ,~~~i=1,\cdots,8~;
 ~~~~~~~
S_n \rightarrow (-1)^n  S_{n}
\label{z2op}
\end{equation}   

\subsection{Tadpole Cancellation}

Let us proceed now and study the anomaly cancellation in this 
unoriented closed and open string model by computing the relevant 
vacuum amplitudes with zero Euler Character. These are the 
Klein Bottle, the Annulus and the Mobius Strip\footnote{For a recent review of
Open Strings and Tadpole cancellation see \cite{sagnotti} and references 
therein.}. 

The amplitude can be expressed in terms of the modified modular 
$f$ functions \cite{bgg}:

\bea
f_{1}^{(m)}(q)&=& q^{-\frac{1}{2}\Delta_{(m)}}\,\,(1-q^m)^{\frac{1}{2}}
\prod_{n=1}^{\infty} (1-q^{\sqrt{m^2 + n^2}})\nonumber\\
f_{2}^{(m)}(q)&=& q^{-\frac{1}{2}\Delta_{(m)}}\,\,(1+q^m)^{\frac{1}{2}}
\prod_{n=1}^{\infty} (1+q^{\sqrt{m^2 + n^2}})\nonumber\\
f_{3}^{(m)}(q)&=& q^{-\frac{1}{2}\Delta'_{(m)}}\,\,
\prod_{n=1}^{\infty} (1+q^{\sqrt{m^2 + (n-\frac{1}{2})^2}})\nonumber\\
f_{4}^{(m)}(q)&=& q^{-\frac{1}{2}\Delta'_{(m)}}\,\,
\prod_{n=1}^{\infty} (1-q^{\sqrt{m^2 + (n-\frac{1}{2})^2}})
\nonumber\\
\eea

where $q=e^{-2\pi t}$, $\Delta_{(m)}$ was given in (\ref{zeropoint}) and 
$\Delta'_{(m)}$ is given by 

\be
\Delta'_{(m)}=-\frac{1}{2\pi^2}\sum_{p=1}^{\infty} (-1)^p
\int^{\infty}_{0} ds \, \, e^{-p^2s-\frac{\pi^2m^2}{s}}.
\ee

Note that in the limit $m\to 0$ these function can be expressed in terms of
standard $\theta$-functions and $\Delta_{(m)}\to -1/12$ and 
$\Delta'_{(m)}\to 1/24$.

Under modular transformation $l=1/t$ they transform as follows:

\be
f_{1}^{(m)}(q)=f_{1}^{(\tilde{m})}(\tilde{q}),\,\,\,\,\,\,\, 
f_{2}^{(m)}(q)=f_{4}^{(\tilde{m})}(\tilde{q}),\,\,\,\,\,\,\,
f_{3}^{(m)}(q)=f_{3}^{(\tilde{m})}(\tilde{q})
\label{modfran}
\ee

where $\tilde{m}=m/l$ and $\tilde{q}=e^{-2\pi l}$.
   
\vskip 0.5 cm   
{\large \it Klein Bottle}\\

The Klein Bottle amplitude in the direct channel is given by the Torus 
amplitude (\ref{torus}) with the $Z_2$ projections (\ref{z2cl}) and leads

\be
Z_{\cal K} =\frac{1}{2}\int \frac{d\tau_2}{\tau_2}\int dp^{+} dp^{-} 
e^{-2\pi\tau_2 p^{+}p^{-}} {\cal K}(\tau_2|m),
\ee

where ${\cal K}$ is given by

\bea
{\cal K}(\tau_2|m)=\frac{(1+q^{\frac{m}{2}})^2}{(1-q^{\frac{m}{2}})^2}
\left(\frac{f_{1}^{(m)}(q)}{f_{1}^{(m)}(q)}\right)^8
\label{AK}
\eea

with $t=2\tau_2$.
We can easily see that the Klein Bottle amplitude vanish by supersymmetry
after an integration over $p^+$ and $p^-$ as it happens for the Torus 
amplitude. In the limit $p^{+}\to 0$ we find 
from the fermionic zero modes $(8-8)$ i.e. 16 supersymmetries. Whereas, from
the bosonic zero modes we get a factor of $1/\tau_2^3$ by integration
over the momentum in the six Neumann directions. We don't see the Dirichlet 
directions due to the fact that we are siting on the point $x^{7,8}=0$.
To extract the ultraviolet behavior of the Klein Bottle amplitude we should 
go to the transverse channel. This is done by modular transformation 
$l=1/t$. As in the closed string case the modular 
transformation will change at the same time $p^+$ and $p^-$ now to 
$p'^{+}=p^{+}/l$ and $p'^{-}=p^-$. The modular transformed amplitude
is\footnote{Here we will just 
quote the results and will not give the details of the computations which 
can be easily done following \cite{takayanagi} and \cite{bgg}.} 

\bea
\tilde{\cal K}(l|\tilde{m})=\frac{(1+e^{-\pi \tilde{m}})^2}
{(1-e^{-\pi\tilde{m}})^2}
\left(\frac{f_{1}^{(\tilde{m})}(\tilde{q})}
{f_{1}^{(\tilde{m})}(\tilde{q})}\right)^8
\label{AKT}
\eea

where $\tilde{m}=\mu p'^+$. Putting every thing together we find 
$\int_{0}^{\infty} dl$ times 
 
\bea
\tilde{Z}_{\cal K}=\frac{1}{2}\int dp'^{+}dp'^{-} 
e^{-\pi p'^{+}p'^{-}}
\frac{(1+e^{-\pi \mu p'^+})^2}{(1-e^{-\pi \mu p'^+})^2}\,\,\,
\left(\frac{f_{1}^{(\tilde{m})}(\tilde{q})}
{f_{1}^{(\tilde{m})}(\tilde{q})}\right)^8.
\label{klein}
\eea

\vskip 0.5 cm
{\large \it Annulus}\\

In this case the trace is over open string spectrum and in order to account 
for the 
internal degrees of freedom (Chan-Paton factors) we associate a multiplicity 
$N$ to each of the string ends.
The direct channel amplitude can be computed easily using the open string 
Hamiltonian (\ref{hop}) leading

\be
Z_{\cal A} =\frac{N^2}{2}\int \frac{d\tau_2}{\tau_2}\int dp^{+} dp^{-} 
e^{-2\pi\tau_2 p^{+}p^{-}} {\cal A}(\tau_2|m),
\label{ZA}
\ee

where ${\cal A}$ is given by 

\be
{\cal A}(\tau_2|m)=\frac{q^{-m}}{(1-q^{m})^2}
\left(\frac{f_{1}^{(m)}(q)}{f_{1}^{(m)}(q)}\right)^8
\label{AA}
\ee

with $t=\tau_2/2$ for the Annulus. As for the Klein Bottle the 
Annulus amplitude also vinishes by integration over the  
$p^\pm$ momentum. Making a modular transformation
to go to the transverse channel $l=1/t$ the momentum $p^+$ and $p^-$
get changed to $p'^{+}=p^{+}/l$ and $p'^{-}=4p^-$. In terms of 
$\tilde{q}$ the amplitude will lead

\bea
\tilde{{\cal A}}(l|\tilde{m})=
\frac{e^{-2\pi \tilde{m}}}{(1-e^{-2\pi\tilde{m}})^2}
\left(\frac{f_{1}^{(\tilde{m})}(\tilde{q})}
{f_{1}^{(\tilde{m})}(\tilde{q})}\right)^8
\label{AAT}
\eea

where as before $\tilde{m}=\mu p'^{+}$. Substituting (\ref{AAT}) in 
(\ref{ZA}) and making the change of variables we find 
$\int_{0}^{\infty} dl$ times

\bea
\tilde{Z}_{\cal A}=\frac{N^2}{8}\int dp'^{+}dp'^{-} 
e^{-\pi p'^{+}p'^{-}}
\frac{e^{-2\pi \mu p'^+}}{(1-e^{-2\pi \mu p'^+})^2}
\left(\frac{f_{1}^{(\tilde{m})}(q)}
{f_{1}^{(\tilde{m})}(q)}\right)^8.
\label{annulus}
\eea

\vskip 0.5 cm
{\large \it Mobius Strip}\\

To compute the Mobius strip amplitude we should make the trace using the 
Hamiltonian (\ref{hop}) over the Hilbert space of open string states 
projected by the $Z_2$ projection (\ref{z2op}). A straightforward  
computation gives

\be
Z_{\cal M} =-\frac{N}{2}\int \frac{d\tau_2}{\tau_2}\int dp^{+} dp^{-} 
e^{-2\pi\tau_2 p^{+}p^{-}} {\cal M}(\tau_2|m),
\label{ZM}
\ee

with $t=\tau_2$ we have 

\bea
{\cal M}(\tau_2|m)=\frac{q^{\frac{m}{2}}}{(1-q^{\frac{m}{2}})^2}\,\,
\left(\frac{f_{1}^{(\frac{m}{2})}(q)f_{3}^{(\frac{m}{2})}(q)}
{f_{1}^{(\frac{m}{2})}(q)f_{3}^{(\frac{m}{2})}(q)}\right)^8.
\label{AM}
\eea

We can now perform a modular transformation 
to go to the transverse channel. In this case we should first take 
$t'=1/t$ and then $l=t'/2$, this will also change the momentum
to $p'^{+}=p^{+}/2l$ and $p'^{-}=2p^-$. The transverse channel 
amplitude will leads

\bea
\tilde{{\cal M}}(l|\tilde{m})=
\frac{e^{-\pi \tilde{m}}}{(1-e^{-\pi \tilde{m}})^2}\,\,
\left(\frac{f_{1}^{(\frac{\tilde{m}}{2})}(\tilde{q})
f_{3}^{(\frac{\tilde{m}}{2})}(\tilde{q})}
{f_{1}^{(\frac{\tilde{m}}{2})}(\tilde{q})
f_{3}^{(\frac{\tilde{m}}{2})}(\tilde{q})}\right)^8,
\label{AMT}
\eea

where as before $\tilde{m}=\mu p'^+$. Substituting \ref{AMT} into
(\ref{ZM}) and making the appropriate change of variables we find 
$\int_{0}^{\infty} dl$ times

\bea
\tilde{Z}_{\cal M}=-\frac{N}{2}\int dp'^{+}dp'^{-} 
e^{-\pi p'^{+}p'^{-}}
\frac{e^{-\pi \mu p'^+}}{(1-e^{-\pi \mu p'^+})^2}
\left(\frac{f_{1}^{(\frac{\tilde{m}}{2})}(\tilde{q})
f_{3}^{(\frac{\tilde{m}}{2})}(\tilde{q})}
{f_{1}^{(\frac{\tilde{m}}{2})}(\tilde{q})
f_{3}^{(\frac{\tilde{m}}{2})}(\tilde{q})}\right)^8.
\label{mobius}
\eea

 
Note that as for the closed string case each of the 
three direct channel amplitudes (\ref{AK}), (\ref{AA}) and (\ref{AM}) is 
zero after the integration over the momentum $p^+$ and
$p^-$, due to the fermionic zero modes reflecting the fact that the model 
is supersymmetric with 
sixteen supersymmetries\footnote{This is the same as in 
the flat background where the three amplitudes give zero separately due to the 
cancellation between the NS-NS and R-R sectors.} i.e. $(8-8)$. However,
in the transverse channel The integration over the $p^+$ and $p^-$ momentum 
should be performed now with
some care due to the appearance of a new singularity\footnote{A similar 
factor was analysed in \cite{AN} contributing $\frac{1}{2\pi \tilde{m}}$ that
was interpreted as a volume factor.} coming from the factor 
$(1-e^{-\pi \mu p'^+})^2$ which obviously can not be reproduced by a 
closed-string amplitude calculation being independent of $l$. 

We can now study the ultraviolet behavior of the three 
amplitudes, in the  
limit $l\to \infty$ the nontrivial contribution comes from the part which is 
independent of $l$ (ground state) and we find

\bea
\tilde{Z}_{\cal K}+\tilde{Z}_{\cal A}+\tilde{Z}_{\cal M} 
&=&\frac{1}{2}\int dp^+ dp^- 
\frac{e^{-\pi p^+ p^-}}{(1-e^{-\pi \mu p^+})^2}
\left[(1+e^{-\pi \mu p^+})^2 + \frac{N^2 e^{-2\pi \mu p^+}}
{4(1+e^{-\pi \mu p^+})^2}-N e^{-\pi \mu p^+}\right]\nonumber\\
&=&\frac{1}{2}\int dp^+ dp^- 
\frac{e^{-\pi p^+ p^-}}{(1-e^{-\pi \mu p^+})^2}
\left[(1+e^{-\pi \mu p^+}) - \frac{N e^{-\pi \mu p^+}}
{2(1+e^{-\pi \mu p^+})}\right]^2.
\eea

It turns out that for $N=8$ the integrand is a regular function of $p^+$,  

\be
\frac{1}{2}\int dp^+ dp^- e^{-\pi p^+ p^-}
\frac{(1-e^{-\pi \mu p^+})^2}{(1+e^{-\pi \mu p^+})^2}
\ee

and gives zero by integrating over $p^+$ and $p^-$. Otherwise the integrand
is singular and the integral will diverge. Therefore, for the Chan-Paton gauge 
group $SO(8)$ there is tadpole cancellation. Note that because we are 
working in the G-S formulation of the theory we are not able to see the 
vanishing of the NS-NS and R-R tadpoles separately but every thing mix-up
in this formalism. In \cite{AN} the authors proposed a way to read off 
the contribution of the R-R states by inserting a projection operator 
$\frac{1-(-1)^{F_L}}{2}$ in the tree and one loop channel amplitudes. We 
have not attempted to do this in the present work.   

\section{Conclusion}

In this paper we have computed the partition function of type IIB string 
theory in plane wave R-R 5-form background using both path integral and 
operator formalisms.
We have shown that the two give the same result and is equal to the 
partition function of type IIB theory in flat background. 
The same observation apply to type IIB in plane wave 
NS-NS and R-R 3-form backgrounds of \cite{rt}. The partition function of the
NW model was computed in \cite{kiritsis} and found to be equal to that of
flat space. It would be very interesting to 
determine the massless vertex operators and compute correlation functions in 
the plane wave R-R background. The one-loop amplitudes will be important 
for the comparison with the non-planar diagrams in the gauge theory. 

We have studied the tadpole cancellation in a model of unoriented 
closed and open strings in plane wave background dual to ${\cal N}=2$
$Sp(N)$ gauge theory with hypermultiplet in the antisymmetric 
representation and four fundamental hypermultiplets. We have seen that there is
tadpole cancellation for SO(8) gauge group. It will be very interesting to 
have a NSR formulation of string theory in plane wave R-R background.

\newpage
{\large Acknowledgments}:\\
I am very greatfull to Edi Gava, Matthias Blau, Nouraddine Chair,
Kumar S. Narain, Jorge G. Russo, George Thompson, Elias Kiritsis
and Jose F. Morales for very illuminating discussions. I would like to 
thank the High Energy Group of the Abdus Salam International Center for 
Theoretical Physics for Hospitality and Grants. This work is partially 
supported by ATN contract: HPRN-CT-2000-00122.



\begin{thebibliography}{99}

\bibitem{blau1}
M.~Blau, J.~Figueroa-O'Farrill, C.~Hull and G.~Papadopoulos,
``A new maximally supersymmetric background of IIB superstring theory,''
JHEP {\bf 0201} (2002) 047, hep-th/0110242.

\bibitem{blau2}
M.~Blau, J.~Figueroa-O'Farrill, C.~Hull and G.~Papadopoulos,
``Penrose limits and maximal supersymmetry,''
Class.\ Quant.\ Grav.\  {\bf 19} (2002) L87, hep-th/0201081;
M.~Blau, J.~Figueroa-O'Farrill and G.~Papadopoulos,
``Penrose limits, supergravity and brane dynamics,'' hep-th/0202111.

\bibitem{m}
R.~R.~Metsaev,
``Type IIB Green-Schwarz superstring in plane wave Ramond-Ramond
background,''
Nucl.\ Phys.\ B {\bf 625} (2002) 70, hep-th/0112044;
R.~R.~Metsaev and A.~A.~Tseytlin,
``Exactly solvable model of superstring in plane wave Ramond-Ramond
background,'' hep-th/0202109.

\bibitem{rt}
J.G.~Russo and A.A.~Tseytlin
``On solvable models of type 2B superstring in NS-NS and R-R wave 
backgrounds'', JHEP {\bf 0204} (2002) 021, hep-th/0202179.

\bibitem{bmn}
D.~Berenstein, J.~M.~Maldacena and H.~Nastase,
``Strings in flat space and pp waves from N = 4 super Yang Mills,''
JHEP {\bf 0204} (2002) 013, hep-th/0202021.


\bibitem{gnbmn}
D.~Berenstein, E.~Gava, J.~M.~Maldacena, K.~S.~Narain and H.~Nastase,
``Open Strings On Plane Waves and Their Yang-Mills Duals,''
hep-th/0203249.

\bibitem{FS}
A.~Fayyazuddin and M.~Spalinski, "Large N
Superconformal Gauge Theories and Supergravity Orientifolds",
Nucl. Phys. {\bf B535} (1998) 219, hep-th/9805096;
O.~Aharony, A.~Fayyazuddin and J.~Maldacena,
"The Large N limit of ${\cal N}=1,2$ from Three-Branes in
F-Theory",  JHEP {\bf 9807}(1998) 013, hep-th/9807159.

\bibitem{tnw}
A.~A.~Tseytlin,
unpublished work.

\bibitem{takayanagi}
T.~Takayanagi,
``Modular invariance of strings on pp-waves with RR-flux,''
hep-th/0206010.

\bibitem{KM}
R.~Kallosh and A.~Morozov,
``Green-Scwarz Action and Loop calculation for Superstrings,''
Int.J.Mod.Phys. {\bf A} 3(1988)1943-1958.

\bibitem{CKR}
B. ~Craps, D. ~Kutasov, G. ~Rajesh,
``String Propagation in the Presence of Cosmological Singularities'',
JHEP {\bf 0206} (2002) 053, hep-th/0205101.

\bibitem{kiritsis}
E. ~Kiritsis, C. ~Kounnas,
``String Propagation in Gravitational Wave Background,''
Phys.Lett. {\bf B} 320:264-272,1994, hep-th/9310202.
E. ~Kiritsis, C. ~Kounnas, D. ~Lust,
``Superstring Gravitational Wave Backgrounds with Space-Time Supersymmetry,''
Phys.Lett. {\bf B} 331:321-329,1994, hep-th/9404114.
E. ~Kiritsis, B. ~Pioline,
``Strings in homogeneous gravitational waves and null holography,''
JHEP {\bf 0208} (2002) 048, hep-th/0204004.

\bibitem{bgg}
O.~Bergman, M.~R.~Gaberdiel and M.~B.~Green,
``D-brane interactions in type IIB plane-wave background,''
hep-th/0205183.

\bibitem{sagnotti}
C. ~Angelantonj and A, ~Sagnotti,
``Open Strings,''
hep-th/0204089.
  
\bibitem{AN}
A.~Sinha, N. V ~Suryanarayana,
``Tadpole Analysis of Orientifolded Plane-Waves,''
hep-th/0209247.

\end{thebibliography}
\end{document}